# Dynamic Janus Metasurfaces in the Visible Spectral Region

*Ping Yu[1], Jianxiong Li[1], Shuang Zhang[2], Zhongwei Jin[3], Gisela Schütz[1], Cheng-Wei Qiu[3,4],\*, Michael Hirscher[1],\*, and Na Liu[1,5],\**

[1]Max Planck Institute for Intelligent Systems, Heisenbergstrasse 3, 70569 Stuttgart, Germany
[2]School of Physics & Astronomy, University of Birmingham, Birmingham B15 2TT, UK.
[3]Department of Electrical and Computer Engineering, National University of Singapore, 4 Engineering Drive 3, Singapore 117583, Singapore
[4]NUS Suzhou Research Institute (NUSRI), Suzhou Industrial Park, Suzhou 215123, China
[5]Kirchhoff Institute for Physics, University of Heidelberg, Im Neuenheimer Feld 227, 69120 Heidelberg, Germany.

Abstract:

Janus monolayers have long been captivated as a popular notion for breaking in-plane and out-of-plane structural symmetry. Originated from chemistry and materials science, the concept of Janus functions have been recently extended to ultrathin metasurfaces by arranging meta-atoms asymmetrically with respect to the propagation or polarization direction of the incident light. However, such metasurfaces are intrinsically static and the information they carry can be straightforwardly decrypted by scanning the incident light directions and polarization states, once the devices are fabricated. In this Letter, we present a dynamic Janus metasurface scheme in the visible spectral region. In each super unit cell, three plasmonic pixels are categorized into two sets. One set contains a magnesium nanorod and a gold nanorod that are orthogonally oriented with respect to each other, working as counter pixels. The other set only contains a magnesium nanorod. The effective pixels on the Janus metasurface can be reversibly regulated by hydrogenation/dehydrogenation of the magnesium nanorods. Such dynamic controllability at visible frequencies allows for flat optical elements with novel functionalities including beam steering, bifocal lensing, holographic encryption, and dual optical function switching.

The god Janus in Roman mythology was seen as two faced, looking to the past and the future, respectively. Inspired by this implication, scientists have named two-faced particles that possess different materials on the opposite sides as "Janus particles".[1, 2] The material anisotropy and directionality endow Janus particles with rich functions and extra controllability,[3-6] allowing for a variety of applications in self-propelled micro-/nanomachines[7, 8] and drug delivery.[9-11] A similar concept has been applied in two-dimensional (2D) materials. Janus monolayers obtained by breaking the structural symmetry of 2D materials can give rise to giant band gap tunability, offering a novel mechanism to tailor the physical properties of electronic nanodevices.[12-18]

In the past several years, the Janus concept has also been attempted in metasurfaces by combining two sets of antennas within a single structural layer. The two sets of antennas can respond differently to different polarizations of light,[19-22] enabling interesting optical function switching. However, such switching capabilities are quite limited and they keenly rely on polarization changes. Most critically, the associated optical information can be easily decrypted through scanning the polarization states of the incident light. To date, the field of metasurfaces has already cultivated into a new realm, where dynamic functionalities are highly desirable for further application advancements. Very recently, beam switching and bifocal lensing have been nicely demonstrated in the mid-infrared spectral region by integration of phase-change material GeSbTe with a metasurface.[23] However, the switching process was not reversible and possessed a narrowband nature due to its design principles.

Here we demonstrate dynamic Janus metasurfaces in the visible spectral region, which offer a wealth of dynamic optical functionalities. Each super unit cell comprises two sets of metal nanorods. In one set, two orthogonally oriented nanorods made of gold (Au) and magnesium (Mg) work as counter pixels. Mg nanorods are adopted here for their remarkable plasmonic reconfigurability in response to hydrogen ($H_2$) and oxygen ($O_2$) at optical frequencies.[24-26] More specifically, Mg can undergo a phase-transition from metal to dielectric upon $H_2$ loading, forming magnesium hydride ($MgH_2$). This transition is reversible through dehydrogenation using $O_2$. As a result, the plasmonic response of an Mg nanorod can be reversibly switched on and off, constituting a dynamic plasmonic pixel. In the other set, only Mg nanorods are utilized to serve as dynamic pixels. With such a triple-pixel scheme, we demonstrate several important dynamic metasurface functionalities, including beam steering, bifocal lensing, dynamic holography, and dual optical function switching at visible frequencies with high fidelity and good reversibility.

Here, Pancharatnam-Berry (PB) phase is utilized to manipulate the wavefront of light, which results from controlling the rotating direction of the electric field of a circularly polarized incident (CP) light by anisotropic scatterers. In this work, right-handed circularly polarized (RCP) incident light is used for all experiments. Figure 1a shows the simulated phase delay and reflectance of the anomalous RCP light in dependence on the Au nanorod orientation angle $\theta$ upon normal incidence of RCP light at 633 nm. The Au nanrods reside on a $SiO_2$/Si substrate. Details of the numerical simulations can be found in Supporting Information. The reflected anomalous RCP wave carries a phase delay of $2\theta$ that spans the entire $2\pi$ range, while its reflectance at resonance remains nearly constant. Such characteristics can also be achieved using Mg nanorods as shown in Figure 1b. By optimizing the dimensions of the Au and Mg nanorods as well as the unit cell size, the reflectance in the two cases can be tuned very close to one another. When a Mg nanorod and a Au nanorod are placed in close proximity ($d$ = 300 nm), the total reflectance becomes highly dependent on their relative orientation angle $\theta$ (see Figure 1c). At approximately $\theta$ = 90°, the anomalous RCP waves reflected from the Au and Mg nanorods achieve a $\pi$ phase shift, giving rise to a reflectance minimum in the far field due to destructive interference.

For creation of the proposed Janus metasurfaces, we take a step further by considering a super unit cell consisting of three elements: orthogonally oriented Au and Mg nanorods (named pixels $P_+$ and $P_-$, respectively), as well as an additional Mg nanorod (named pixel P), as shown in Figure 1d. Before hydrogenation, the net function of such a super unit cell is governed by P, as $P_+$ and $P_-$ cancel each other in reflectance. In other words, before hydrogenation the effective pixel of the super unit cell is $\mathbf{P_e}$ = P. After hydrogenation, Mg is transformed into $MgH_2$. The net function of the super unit cell is therefore only governed by $P_+$, i.e. $\mathbf{P_e}$ = $P_+$. Upon $O_2$ exposure, $MgH_2$ can be transformed back to Mg and thus $\mathbf{P_e}$ = P again. Therefore, the effective pixels on such a Janus metasurface can be reversibly regulated using $H_2$ and $O_2$, independent on the helicity of the incident light. This scheme opens a unique pathway to endow optical metasurfaces with rich dynamic functionalities in the optical spectral region.

First, we demonstrate dynamic beam steering using a Janus metasurface. Eight phase levels are used for wavefront shaping. As depicted in Figure 2a, in each super unit cell $P_+$ and $P_-$ are orthogonally arranged with respect to each other, while $P_+$ and P are geometrically engineered on the metasurface for obtaining the acquired phase profiles to generate anomalous reflection of RCP waves along two opposite propagation directions, respectively, at normal incidence of RCP light (633 nm). Figure 2b shows the simulated electric field distributions of the reflected anomalous RCP light upon $H_2$ and $O_2$ loading, respectively. Upon $H_2$ exposure, the reflected beam is steered to the left, whereas after $O_2$ exposure the reflected beam is steered to the right. To experimentally prove the dynamic beam steering, a metasurface has been fabricated on a $SiO_2$/Si substrate using multi-step electron-beam lithography (see Supporting Information).

Figure 2c presents a scanning electron microscopy (SEM) image of the metasurface, in which the Au and Mg nanorods can be easily identified due to their clear material contrast. The metasurface sample is placed in a homemade gas cell and positioned along an optical path for optical characterizations (see the corresponding optical setup in Figure S1). The experimental results as presented in Figure 2d, confirm the beam steering effects upon $H_2$ and $O_2$ loading. The white circles in the snapshot images indicate the location of the zero-order reflected light. The hydrogenation and dehydrogenation processes take approximately 30 seconds and 2 hours to complete at room temperature, respectively. The switching speed can be significantly enhanced using elevated temperatures (see Movie 2) or employing Mg-Ni alloys with a Ta intermediate layer.[27] A video that records the dynamic and reversible beam steering process can be found in Movie S1.

Conventional multiplexed metasurfaces allow for generation of interchangeable holograms controlled by the polarization of the incident light. However, the holographic images can be decrypted easily by scanning the polarization states of the incident light. This impedes the application of metasurface holography in data security and optical information encryption. Our Janus metasurface provides a solution to solve these issues. As illustrated in Figure 3a, each super unit cell (600 nm × 600 nm) comprises three pixels $P_+$, $P_-$ and P. P and $P_+$ are adopted to yield the required phase profiles for generating two independent holographic images 'Y' and 'N', respectively. The phase-only holograms are designed based on Gerchberg-Saxton algorithm. The SEM image of the fabricated metasurface on a $SiO_2$/Si substrate is presented in Figure 3b. The corresponding optical setup can be found in Figure S2. As shown by the experimental results in Figure 3c, before hydrogenation 'Y' is observed on the screen upon illumination of RCP light. After $H_2$ exposure, 'Y' vanishes, whereas 'N' comes into existence. After $O_2$ exposure, 'N' disappears and 'Y' occurs again. Consequently, the two holographic images, containing different information can be independently reconstructed upon $H_2$ and $O_2$ exposures, respectively, while the helicity of the incident light remains unchanged. In this regard, 'Y' and 'N' can appear at the same spatial location alternatively. The encrypted information ('N' in this case) cannot be decrypted by scanning the polarization states of the incident light. $H_2$ is the only decoding key. In addition, the gas operation step can be included as an extra key to further secure the optical information. Pre-knowledge of the exact step is crucially required to obtain the correct information, as even opposite messages can be delivered at different gas operation steps (see Figure 3c). Our scheme therefore suggests promising applications for high-level anticounterfeiting and optical information encryption. It is noteworthy that the residual of 'N' before hydrogenation is due to an imperfect reflectance cancellation from the two orthogonal Au and Mg nanorods (see Figure 1c). This can be avoided by a careful dimension tuning of the Au (or Mg) nanorod for achieving a perfect reflectance cancellation. A video that records the dynamic hologram switching process can be found in Movie S2. To accelerate the switching speed, an elevated temperature of 80°C is utilized so that the entire switching (hydrogenation and dehydrogenation) process takes only about 3 minutes. To prove the broadband nature of our metasurface performance, Figure S3 shows the holographic images captured at wavelength of 520 nm. The switchable holograms achieved under different light polarizations can be found in Figure S4.

Next, we demonstrate a dynamic bifocal cylindrical metasurface lens as depicted in Figure 4a. Pixels P and $P_+$ are adopted to obtain the acquired phase profiles for achieving two distinct focal lengths in transmission with $f_1 = 200$ μm and $f_2 = 100$ μm, respectively, for RCP light incidence at 633 nm. It should be noted that the helicity of the transmitted anomalous light is LCP in this case. The spatial phase profile along the *x* direction, $\varphi_x$, are calculated using the equation

$$\varphi_x = \frac{k}{2}\left(\sqrt{f^2 + x^2} - f\right),$$

where *k* and *f* represent the free space wavevector and the desired focal length, respectively. *x* denotes the *x*-position of the nanorod.

Full-field simulation results in Figure 4b reveal that the converted LCP light can be focused at the predesigned *z*-positions (corresponding to $f_1$ = 200 μm and $f_2$ = 100 μm) before and after hydrogenation, respectively. We define the position of the metalens in the optical path as *z* = 0 so that the different focal lengths can directly correspond to the *z* values of the focal planes. Figure 4c presents the SEM image of the metalens fabricated on a $SiO_2$ substrate. Details of the optical setup can be found in Figure S5. The snapshot images in the left column of Figure 4d reveal the beam shapes at the representative *z*-positions from the metalens before hydrogenation. The metalens creates a focal line at *z* = 239 μm, as highlighted by the green frame. After hydrogenation, a new focal line occurs at *z* = 138 μm, as indicated by the green frame in the right column of Figure 4d. The two measured focal lengths are slightly larger than the theoretical values due to the presence of the gas cell window, which enlarges the overall focal lengths. The bifocal functionality of the metalens can be nicely visualized through operations of $H_2$ and $O_2$ (see Movie S3).

Finally, we demonstrate that our scheme also allows for dynamic switching between different metasurface functions, taking a further step towards realization of flat and compact optical elements. Figure 5a illustrates a dynamic metasurface with dual functions as a proof of concept. P and $P_+$ are adopted to obtain the acquired phase profiles to achieve anomalous refraction and beam focusing (*f* =100 μm), respectively. The SEM image of the metasurface is shown in Figure 5b. Two optical setups are employed to characterize the anomalous beam refraction and beam focusing performance, respectively. As shown in Figure 5c, dynamic switching between anomalous refraction and beam focusing is successfully visualized. Our design scheme can be easily extended to realize dynamic switching between other targeted metasurface functions as well.

In conclusion, we have demonstrated Janus metasurfaces based on a multiple plasmonic pixel scheme involving dynamic plasmonic elements and selective functional cancelation between the elements. Such a scheme enables advanced dynamic optical functions in the visible spectral range. Dynamic beam steering, bifocal lensing, holographic encryption, and dual optical function switching have been demonstrated successfully. Our work features great potentials to achieve diversified optical functions, while keeping individual functions highly independent. The dynamic operation scheme can be easily extended to other wavelength regions as well. This work will shed light on designing and implementing a new generation of compact, secure, and multi-tasking optical nanodevices using metasurfaces. In addition, the knowhow gained from our work will provide insights into smart glass and switchable mirror applications.[28]

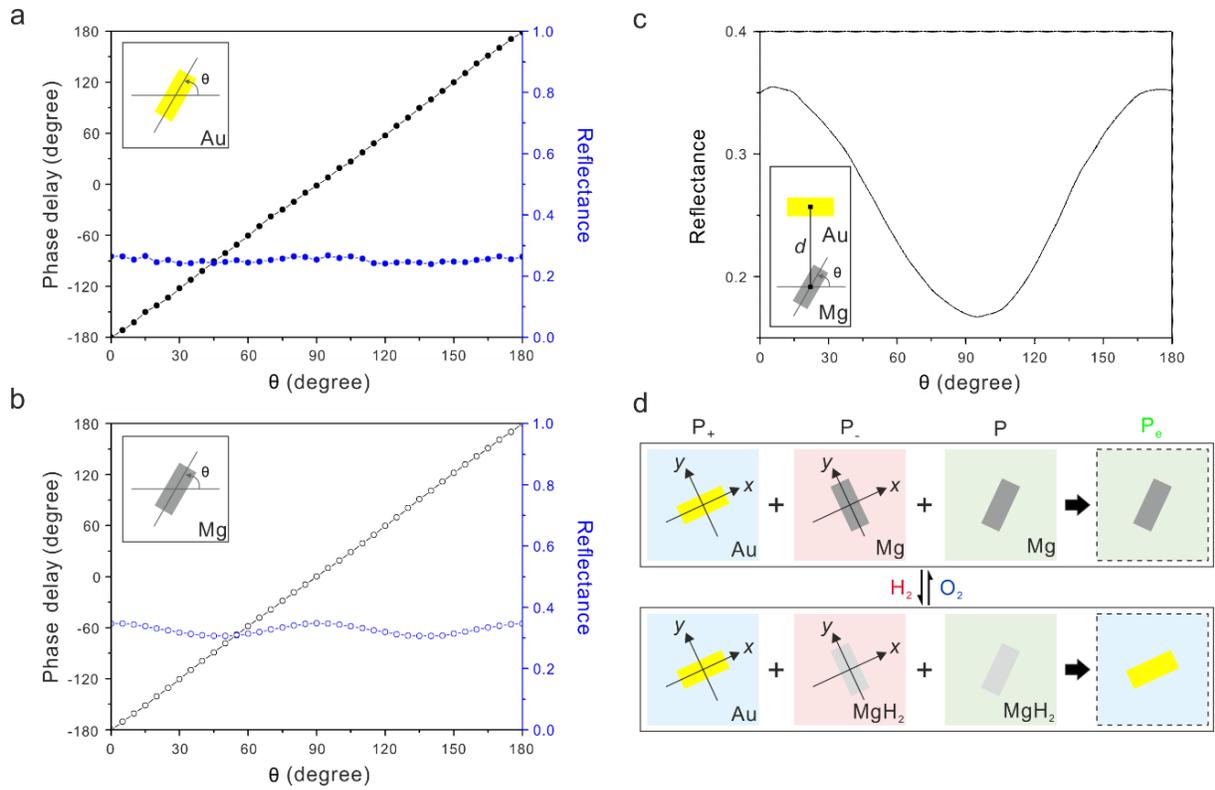

**Figure 1.** Working principle of the dynamic Janus metasurface. Simulated phase delay and reflectance of the anomalous RCP light in dependence on the orientation angle $\theta$ of the Au (a) or Mg (b) nanorod under normal incidence of RCP light at 633 nm. (c) Reflectance of the anomalous RCP light in dependence on the relative angle $\theta$ between the Au and Mg nanorods positioned within a distance $d$ = 300 nm. (d) In a super unit cell, three are three pixels that are categorized in two sets. In one set, Au nanorod ($P_+$) and Mg nanorod ($P_-$) keep orthogonal to one another. In the other set, there exists only one Mg nanorod (P). After hydrogenation (10% $H_2$) and dehydrogenation (20% $O_2$), the effective pixel ($P_e$) corresponds to $P_+$ and P, respectively.

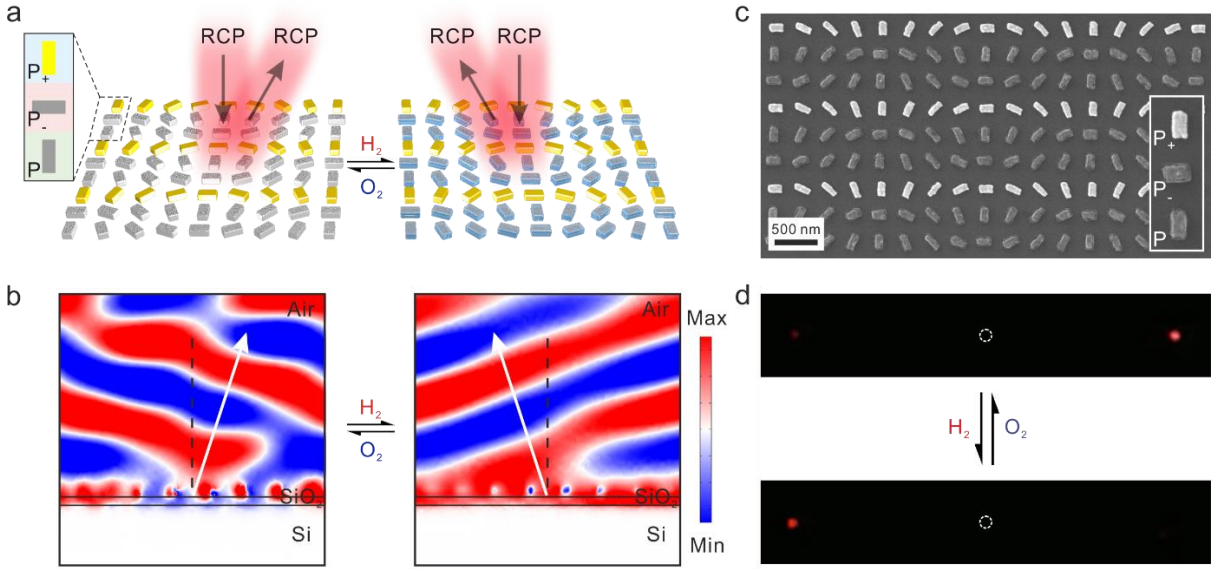

**Figure 2.** Dynamic beam steering. (a) $P_+$ and P are designed to obtain the acquired phase profiles for steering the reflected anomalous RCP light towards two opposite directions under normal incidence of RCP light at 633 nm. The metasurface steers the direction of the light beam upon $H_2$ and $O_2$ exposures accordingly. (b) Simulated electric field distributions of the reflected anomalous RCP light. The nanorods reside on a $SiO_2$ (100 nm)/Si substrate in simulations. (c) Overview SEM image of the plasmonic Janus metasurface. Inset: enlarged SEM image of a super unit cell (300 nm × 900 nm). (d) Representative snapshots of the reflected anomalous light after hydrogenation and dehydrogenation, respectively. The central point in each plot is indicated by a white circle, highlighting the location of the zero-order reflected light.

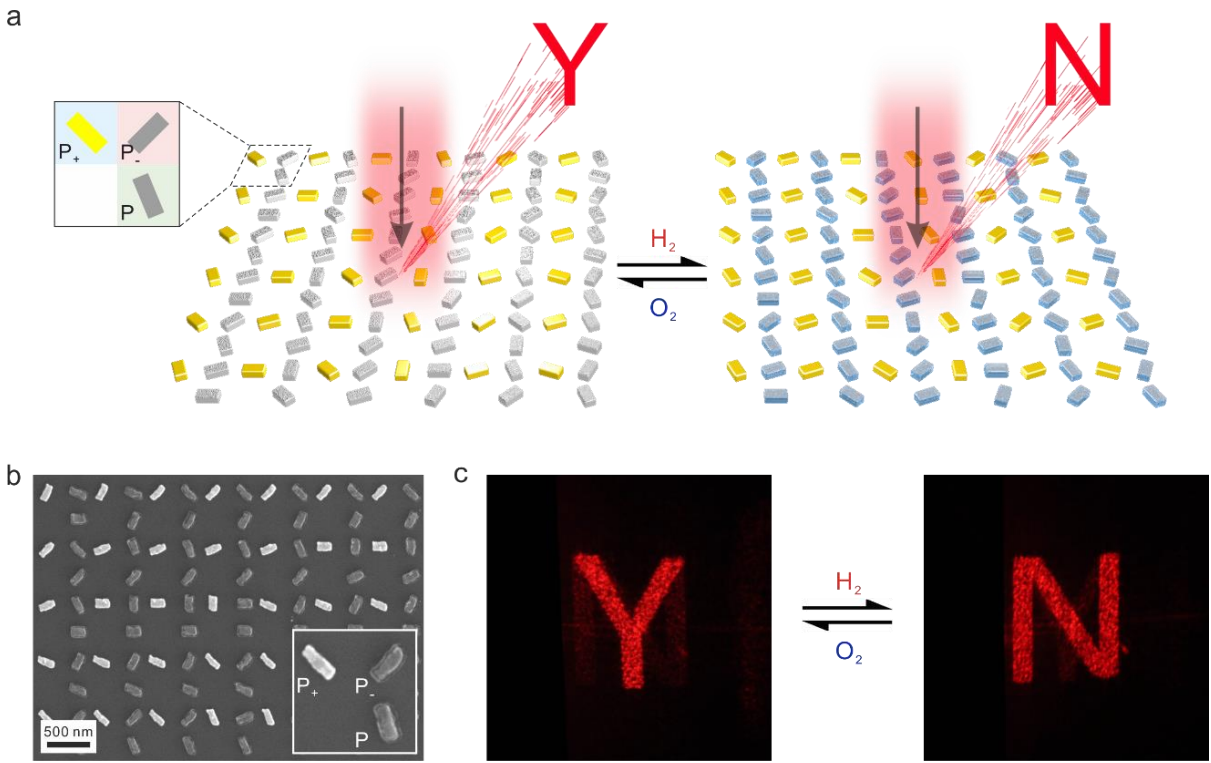

**Figure 3.** Holographic encryption. (a) Two holographic patterns 'Y' and 'N' reconstructed from two independent phase profiles generated using P and P$_+$, respectively. (b) Overview SEM image of the metasurface. Inset: enlarged SEM image of a super unit cell (600 nm × 600 nm). (d) Representative snapshots of the holographic images after hydrogenation and dehydrogenation, respectively. Holographic patterns 'Y' and 'N' can appear at the same spatial location alternatively.

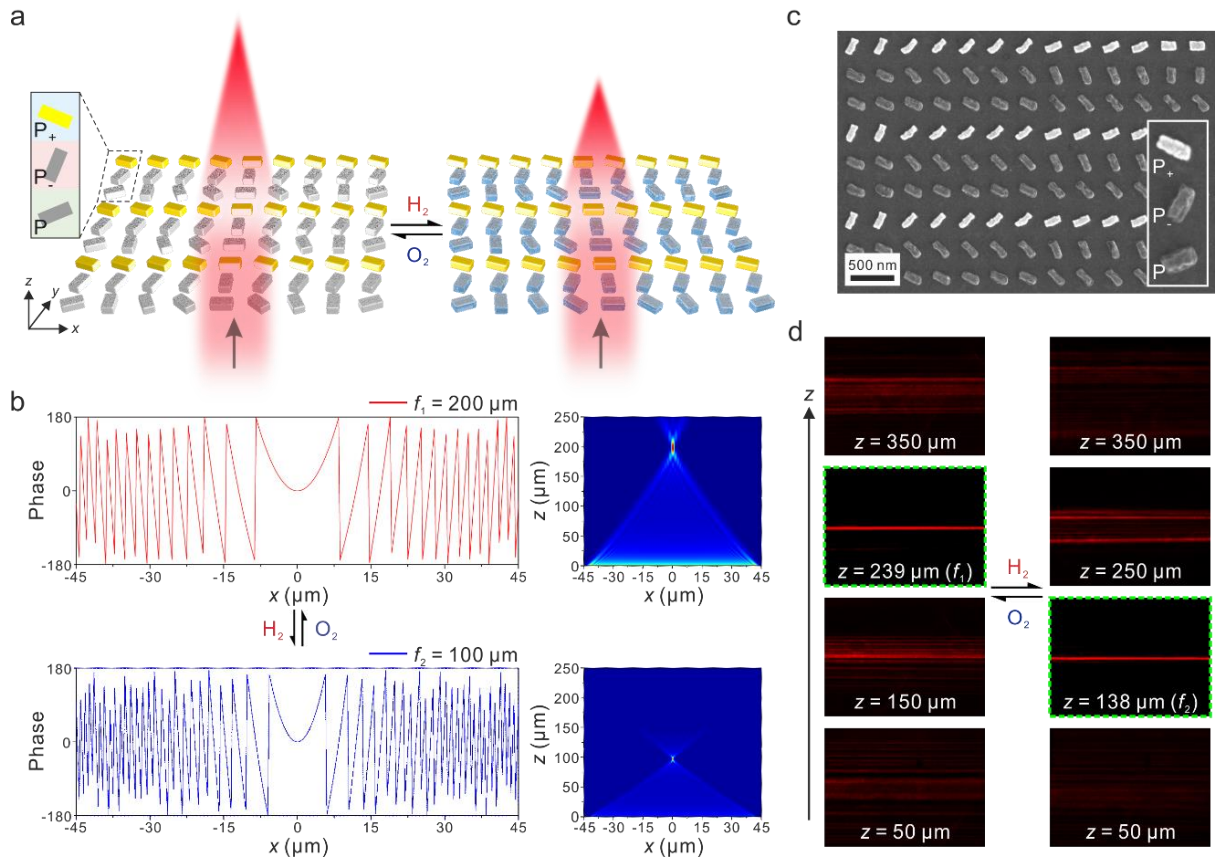

**Figure 4.** Dynamic bifocal metalens. (a) P and P$_+$ are used to obtain the acquired phase profiles with focal lengths of $f_1$ = 200 µm and $f_2$ = 100 µm, respectively. (b) The simulated phase discontinuity profiles and the corresponding full-field intensities of the metalens before and after hydrogenation, respectively. (c) Overview SEM image of the metalens. Inset: enlarged SEM image of a unit cell (300 nm × 900 nm). (d) Representative snapshots in the *x-y* plane at different *z* distances from the metalens after H$_2$ and O$_2$ exposures, respectively. The focusing effects are observed at *z* = 239 µm and *z* = 138 µm before and after hydrogenation, respectively, as indicated by the green frames.

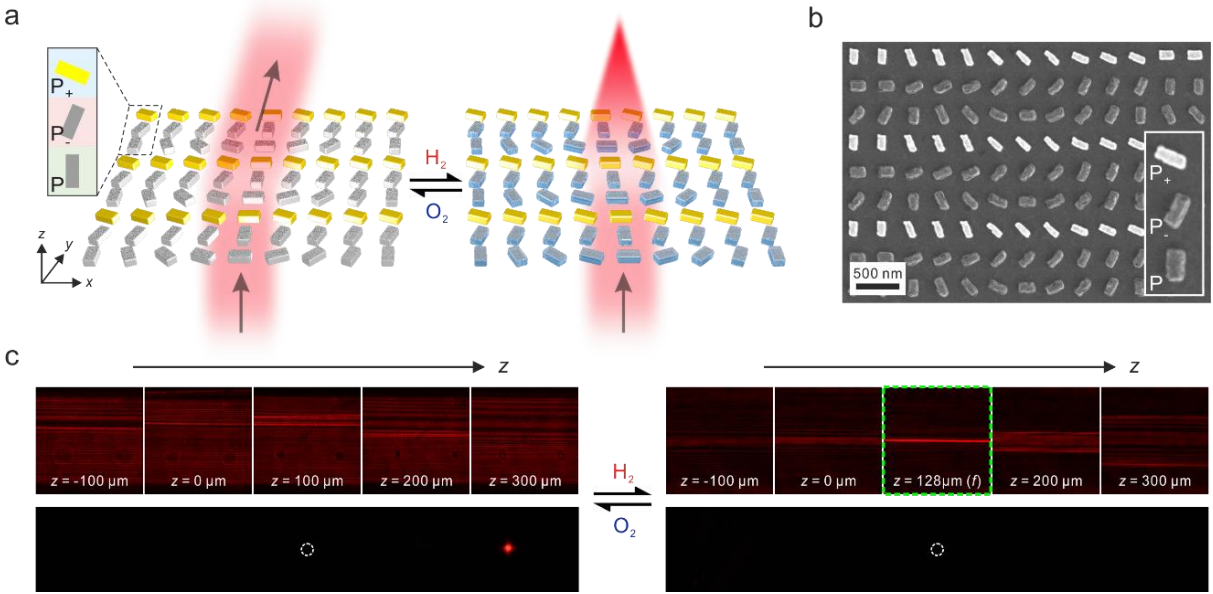

**Figure 5.** Dynamic switching between two independent optical functions. (a) P and P$_+$ are designed to achieve anomalous refraction and beam focusing at $f$ = 100 μm, respectively. (b) Overview SEM image of the dual-function metasurface. Inset: enlarged SEM image of a super unit cell (300 nm × 900 nm). (c) Representative snapshots in the *x-y* plane at different *z* distances from the metasurface and the snapshots of the refracted anomalous light before and after hydrogenation, respectively. The position of the metasurface is defined as $z$ = 0. The white circles in the plots indicate the location of the transmitted ordinary light.

ASSOCIATED CONTENT

**Supporting Information**.

Details of the numerical simulations, sample fabrication and optical setups. The broadband nature of the Janus metasurfaces and the switchable holograms under different light polarizations. The following files are available free of charge.

Supporting information (word)

Movie S1 (AVI)

Movie S2 (AVI)

Movie S3 (AVI)

AUTHOR INFORMATION

Corresponding Author

*Email: chengwei.qiu@nus.edu.sg

*Email: hirscher@is.mpg.de

*Email: na.liu@kip.uni-heidelberg.de

Author Contributions

The manuscript was written through contributions of all authors. All authors have given approval to the final version of the manuscript.

Funding Sources

This project was supported by the Sofja Kovalevskaja grant from the Alexander von Humboldt-Foundation, the European Research Council (ERC Dynamic Nano and ERC Topological) grants, and the Grassroots project from the Max-Planck Institute for Intelligent Systems.

Notes

The authors declare no competing financial interest.

ACKNOWLEDGMENT

We gratefully acknowledge the generous support by the Max-Planck Institute for Solid State Research for the usage of clean room facilities.

REFERENCES

1.	de Gennes, P. G. *Science* **1992**, 256, 495.

2.	Casagrande, C. *CR Acad. Sci. Ser. II* **1998**, 306, 1423-1425.

3.	Hu, S.-H.; Gao, X. *J. Am. Chem. Soc.* **2010**, 132, 7234-7237.

4.	Wu Liz, Y.; Ross Benjamin, M.; Hong, S.; Lee Luke, P. *Small* **2010**, 6, 503-507.

5.	Jiang, J.; Gu, H.; Shao, H.; Devlin, E.; Papaefthymiou Georgia, C.; Ying Jackie, Y. *Adv. Mater.* **2008**, 20, 4403-4407.

6.	Huck, C.; Vogt, J.; Sendner, M.; Hengstler, D.; Neubrech, F.; Pucci, A. *ACS Photonics* **2015**, 2, 1489-1497.

7.	Dong, R.; Zhang, Q.; Gao, W.; Pei, A.; Ren, B. *ACS Nano* **2016**, 10, 839-844.

8.	Soler, L.; Magdanz, V.; Fomin, V. M.; Sanchez, S.; Schmidt, O. G. *Acs Nano* **2013**, 7, 9611-9620.

9.	Hwang, S.; Lahann, J. *Macromol. Rapid Commun.* **2012**, 33, 1178-1183.

10.	Garbuzenko, O. B.; Winkler, J.; Tomassone, M. S.; Minko, T. *Langmuir* **2014**, 30, 12941-12949.

11.	Wang, F.; Pauletti, G. M.; Wang, J.; Zhang, J.; Ewing, R. C.; Wang, Y.; Shi, D. *Adv. Mater.* **2013**, 25, 3485-3489.

12.	Zhang, Y.; Tang, T.-T.; Girit, C.; Hao, Z.; Martin, M. C.; Zettl, A.; Crommie, M. F.; Shen, Y. R.; Wang, F. *Nature* **2009**, 459, 820-823.

13.	Shimazaki, Y.; Yamamoto, M.; Borzenets, I. V.; Watanabe, K.; Taniguchi, T.; Tarucha, S. *Nat. Phys.* **2015**, 11, 1032-1036.

14.	Hunt, B.; Sanchez-Yamagishi, J. D.; Young, A. F.; Yankowitz, M.; LeRoy, B. J.; Watanabe, K.; Taniguchi, T.; Moon, P.; Koshino, M.; Jarillo-Herrero, P.; Ashoori, R. C. *Science* **2013**, 340, 1427-1430.

15.	Ye, Z.; Cao, T.; O'Brien, K.; Zhu, H.; Yin, X.; Wang, Y.; Louie, S. G.; Zhang, X. *Nature* **2014**, 513, 214-218.

16.	Yuan, H.; Bahramy, M. S.; Morimoto, K.; Wu, S.; Nomura, K.; Yang, B.-J.; Shimotani, H.; Suzuki, R.; Toh, M.; Kloc, C.; Xu, X.; Arita, R.; Nagaosa, N.; Iwasa, Y. *Nat. Phys.* **2013**, 9, 563-569.

17. Wu, S.; Ross, J. S.; Liu, G.-B.; Aivazian, G.; Jones, A.; Fei, Z.; Zhu, W.; Xiao, D.; Yao, W.; Cobden, D.; Xu, X. *Nat. Phys.* **2013**, 9, 149-153.

18. Lu, A.-Y.; Zhu, H.; Xiao, J.; Chuu, C.-P.; Han, Y.; Chiu, M.-H.; Cheng, C.-C.; Yang, C.-W.; Wei, K.-H.; Yang, Y.; Wang, Y.; Sokaras, D.; Nordlund, D.; Yang, P.; Muller, D. A.; Chou, M.-Y.; Zhang, X.; Li, L.-J. *Nat. Nanotechnol.* **2017**, 12, 744-749.

19. Montelongo, Y.; Tenorio-Pearl, J. O.; Milne, W. I.; Wilkinson, T. D. *Nano Lett.* **2014**, 14, 294-298.

20. Balthasar Mueller, J. P.; Rubin, N. A.; Devlin, R. C.; Groever, B.; Capasso, F. *Phys. Rev. Lett.* **2017**, 118, 113901.

21. Wen, D.; Yue, F.; Li, G.; Zheng, G.; Chan, K.; Chen, S.; Chen, M.; Li, K. F.; Wong, P. W.; Cheah, K. W.; Pun, E. Y.; Zhang, S.; Chen, X. *Nat. Commun.* **2015**, 6, 8241.

22. Chen, X.; Huang, L.; Mühlenbernd, H.; Li, G.; Bai, B.; Tan, Q.; Jin, G.; Qiu, C.-W.; Zhang, S.; Zentgraf, T. *Nat. Commun.* **2012**, 3, 1198.

23. Yin, X.; Steinle, T.; Huang, L.; Taubner, T.; Wuttig, M.; Zentgraf, T.; Giessen, H. *Light Sci. Appl.* **2017**, 6, e17016.

24. Duan, X.; Kamin, S.; Sterl, F.; Giessen, H.; Liu, N. *Nano Lett.* **2016**, 16, 1462-1466.

25. Duan, X.; Kamin, S.; Liu, N. *Nat. Commun.* **2017**, 8, 14606.

26. Sterl, F.; Strohfeldt, N.; Walter, R.; Griessen, R.; Tittl, A.; Giessen, H. *Nano Lett.* **2015**, 15, 7949-7955.

27. Yamada, Y.; Miura, M.; Tajima, K.; Okada, M.; Yoshimura, K. *Solar Energy Materials and Solar Cells* **2014**, 125, 133-137.

28. Yoshimura, K.; Langhammer, C.; Dam, B. *MRS Bulletin* **2013**, 38, 495-503.

**TOC graphic**

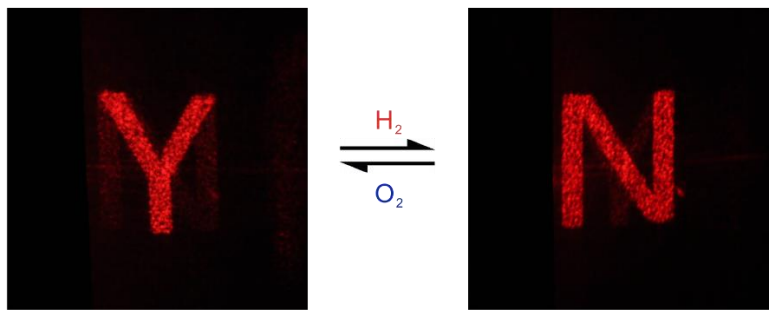